\documentstyle[jkas]{article}

\beginpage{1}
\endpage{4}
\year{2010}\volume{43}\month{February} 

\runningauthor {H.S. PARK} 
\runningtitle{PHOTOMETRY OF M86 GLOBULAR CLUSTERS}

\month{February} \year{2010} \volume{43} \issueno{4}
\beginpage{125}\endpage{130}
\date{Received December 1, 2009; Accepted January 17, 2010}

\begin{document}
\title{WASHINGTON PHOTOMETRY OF THE GLOBULAR CLUSTERS IN THE VIRGO GIANT ELLIPTICAL GALAXY M86} 
\author{Hong Soo Park$^{1}$}
\address{$^1$ Department of Physics and Astronomy, Seoul National University,
Seoul 151-742, Korea\\ {\it E-mail : hspark@astro.snu.ac.kr }}

\address{\normalsize{\it (Received May 00, 2012; Accepted May 00, 2012)}}
\offprints{H. S. Park} 
\abstract{
We present a photometric study of the globular clusters (GCs) 
in the Virgo giant elliptical galaxy M86
based on Washington $CT_1$ images.
The colors of the GCs in M86 show a bimodal distribution
with a blue peak at $(C-T_1)=1.30$ %
 and a red peak at $(C-T_1)=1.72$. %
The spatial distribution of the red GCs is elongated similarly to that of the stellar halo,
while that of the blue GCs is roughly circular.
The radial number density profile of the blue GCs is more extended than that of the red GCs.
The radial number density profile of the red GCs is consistent with the surface brightness profile of the M86 stellar halo.
The GC system has a negative radial color gradient, which is mainly
due to the number ratio of the blue GCs to the red GCs increasing as galactocentric radius increase.
The bright blue GCs in the outer region of M86 show a blue tilt: 
   the brighter they are, the redder their mean colors get.
These results are discussed in comparison with other Virgo giant elliptical galaxies.  
}

\keywords{galaxies: clusters --- galaxies: individual (M86, NGC 4406) ---
galaxies: photometry --- galaxies: star clusters
}
\maketitle

\section{INTRODUCTION}

Thousands of globular clusters (GCs) are found in a giant elliptical galaxy (gE)  and 
  they are found to be located from the center to the outer halo of their host galaxy \citep{lee03,bro06}.
  Therefore, the GCs in gEs are a powerful tool to study the structure and evolution of 
  the GC system itself as well as their host galaxy.
  
M86 (NGC 4406, VCC 881) is a famous gE 
  located close to the center region of the Virgo galaxy cluster in the sky
  and is known to be infalling to the Virgo center
  with the relative velocity of about 1300 %
  km s$^{-1}$. %
It is also known early to be abundant with GCs \citep{han77}. 
There are several photometric studies for the GCs in M86
in the literature.
  \citet{coh88} presented $gri$ photometry of GCs at $R<7\arcmin$ in M86 obtained at the Hale 5 m telescope.
  She found that the GC system of M86 is similarly extended as the stellar light, and 
  that there is no detectable radial color gradient. %
  
From HST/WFPC2 $VI$ observation of the central region
of M86, 
  \citet{nei99} reported that  the color distribution of the GCs in M86 shows a single peak.
  However, \citet{kun01} and \citet{lar01} showed later,
  using the same data as used in HST/WFPC2 by \citet{nei99},   that the color distribution of the GCs in M86 is bimodal. 
Later \citet{pen06} confirmed, from HST/ACS $gz$ observation, that the color distribution of M86 GCs is bimodal.

 \citet{jor07} and \citet{vil10} derived a luminosity function of M86 GCs from the same HST/ACS images, 
 and obtained a similar Gaussian peak value at $g$-band,
   $\mu_g=23.950\pm0.097$ and $23.887\pm0.087$, respectively. 
  \citet{pen08} estimated the specific frequency of M86 GCs, $S_N=2.57\pm0.12$, from the HST/ACS images. 
On the other hand, 
  \citet{rho04} investigated 
  M86 GCs  
  using the wide field ($36\arcmin\times36\arcmin$) 
  $BVR$ images obtained using the KPNO 4 m Mayall telescope.
  They found again that M86 GCs show a bimodal color distribution and    that they show a modest negative color gradient with galactocentric radius. 
They derived a specific frequency of $S_N=3.5\pm0.5$,
which is larger than the value derived from the central region by \citet{pen08}.
  
However, previous studies based on HST data covered only a small central region of M86 and those based on ground-based data did not investigate the properties of sub-populations in the GC system and the spatial distribution of the GCs in M86.
 Here we investigate the detailed properties of the M86 GC system %
  using Washington photometry derived from deep and wide CCD images.
The Washington filter system with a wide bandwidth is known to be very sensitive to measuring the metallicity of the GCs \citep{gei90} so that it is useful for studying the properties of sub-populations of the GCs. 
We adopt a distance to M86, derived from the surface brightness fluctuation method by \citet{mei07},
16.9 Mpc ($(m-M)_0 =31.13\pm0.07$).
At this distance one arcsec corresponds to a linear scale of 81 pc.
The basic information of M86 is listed in Table \ref{tab-info}.

This paper is composed as follows. 
Section 2 describes observation and data reduction. 
In \S 3, we present the color-magnitude diagram and color distribution of
the M86 GCs.
We investigate the spatial distribution of the GC system as well as the radial variation of number densities and colors of the GCs.
In \S 4,  we discuss our results and their implication in comparison with 
  other Virgo gE studies. 
Primary results are summarized in \S 5.

\begin{deluxetable}{lcc}
\tablecaption{Basic information of M86\label{tab-info}}
\tablewidth{0pt}
\tablehead{
\colhead{Parameters} & \colhead{Values } & \colhead{References} }
\startdata
R.A., Decl. (J2000) & $12^h~26^m~11.743^s$,  +12$\arcdeg$ 56$\arcmin$ 46.40$\arcsec$ & 1 \\
Total magnitudes              & $V^T=8.90\pm0.05$, $B^T=9.83\pm0.05$ & 2\\
Foreground reddening, $E(B-V)$       & 0.030  & 3\\
Distance, $d$ &  16.86 Mpc ($(m-M)_0=31.13\pm0.07$)  & 4 \\
Systemic radial velocity, $v_p$      & $-244\pm5$ km s$^{-1}$ & 5\\
Effective radius, $R_{eff}$       & 3.59 arcmin ($C$), 3.14 arcmin ($T_1$) & 6\\
Effective ellipticity, $\epsilon_{eff}$  & 0.36 ($C$), 0.33 ($T_1$) & 6 \\
Effective P.A., $\Theta_{eff}$  & 119 deg ($C, T_1$)  & 6 \\
Standard radius,   $R_{25}$   & 4.88 arcmin ($C$), 9.64 arcmin ($T_1$) & 6\\
Standard ellipticity, $\epsilon_{25}$  & 0.40 ($C, T_1$) & 6 \\
Standard P.A., $\Theta_{25}$   &  124 deg ($C, T_1$) & 6 \\
\enddata
\tablerefs{(1) NASA Extragalactic Database; 
(2) \citet{dev91}; (3) \citet{sch98}; (4) \citet{mei07}; (5) \citet{smi00};
(6) This study.
}
\end{deluxetable}

\section{OBSERVATION AND DATA REDUCTION}

\subsection{Observations}

The images of M86 were obtained with the $2048 \times 2048$ pixels CCD camera
  (the pixel scale is 0.47 arcsec pixel$^{-1}$)
  at the KPNO 4 m telescope
  on the photometric night of April 9, 1997.
These images were taken with the Washington $C$ and Kron-Cousins $R$ filters.
  The $R$ filter  was used instead of the Washington $T_1$ filter, 
  as done in most recent Washington studies \citep{gei96, lee98, har04, dir05, bas06, for07, lee08},
  because the sensitivity of the Kron-Cousins $R$ filter is about three times larger than  that of the $T_1$ filter,
  while the effective wavelengths of two filters are very similar 
  as $R=T_1 + 0.003-0.017 (C-T_1 )$ with rms=0.02 \citep{gei96}.
  Hereafter we call this $R$ filter as $T_1$ filter in the rest of this paper.

Table \ref{tab-obs} shows the observation log.
  The size of the field of view is $16\arcmin \times 16\arcmin$, and
  the seeing range is from $1\arcsec$.0 to $1\arcsec.2$.
  Exposure times are 60 s and $4 \times 1500$ s for $C$,
  and 30 s and $2 \times 1000$ s for $T_1$.
Several Washington standard fields in \citet{gei96} were also observed during the observing run.

\begin{figure}[!t]
\epsscale{1.00}
\plotone{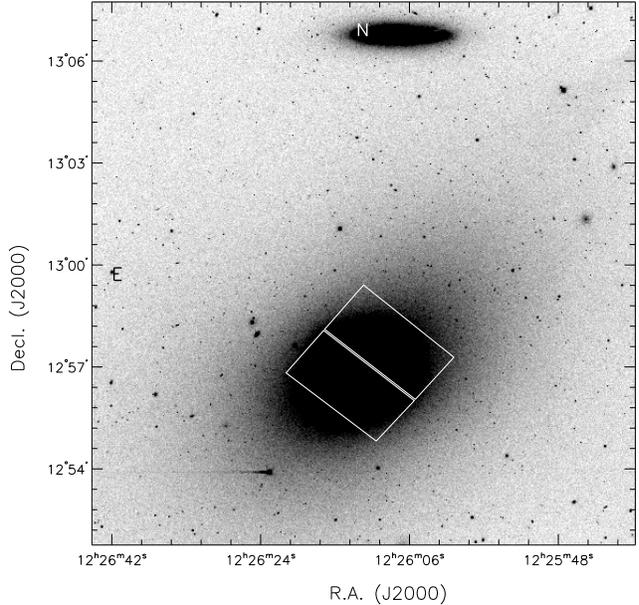}
\caption{
A grey scale map of $T_1$ image of the M86 field taken with KPNO 4 m telescope. 
The field of view is $16\arcmin\times16\arcmin$.
The boxes represent the HST/ACS fields.
The galaxy in the north is NGC 4402, which is an edge-on galaxy of Sb type.
\label{fig-image}}
\end{figure}

\begin{deluxetable}{cccccc}
\tablecaption{Observation log \label{tab-obs}}
\tablewidth{0pt}
\tablehead{
\colhead{Target} & \colhead{Filter} & \colhead{T(exp)} & \colhead{Airmass} & \colhead{Seeing} & \colhead{Date} \\
\colhead{} & \colhead{} & \colhead{(sec)} & \colhead{} & \colhead{($\arcsec$)} & \colhead{(UT)} 
}
\startdata
M86 & C & 60 & 1.12 & $1.13$ & 1997 Apr 9 \\
M86 & R & 30 & 1.06 & $0.99$ & 1997 Apr 9 \\
M86 & C & $4 \times 1500$ & 1.06 & $1.13$ & 1997 Apr 9 \\
M86 & R & $2 \times 1000$ & 1.21 & $1.22$ & 1997 Apr 9 \\
\enddata
\end{deluxetable}

\subsection{Point Source Photometry}

Using the IRAF software, each frame was  
  bias-subtracted and  
  flat-fielded. 
  Then the long exposure images of two $T_1$-bands and four $C$-bands were combined with average and median procedure, respectively.
Figure \ref{fig-image} displays a $T_1$ image of M86 taken with the short exposure.

\begin{figure}[!t]
\epsscale{0.9}
\plotone{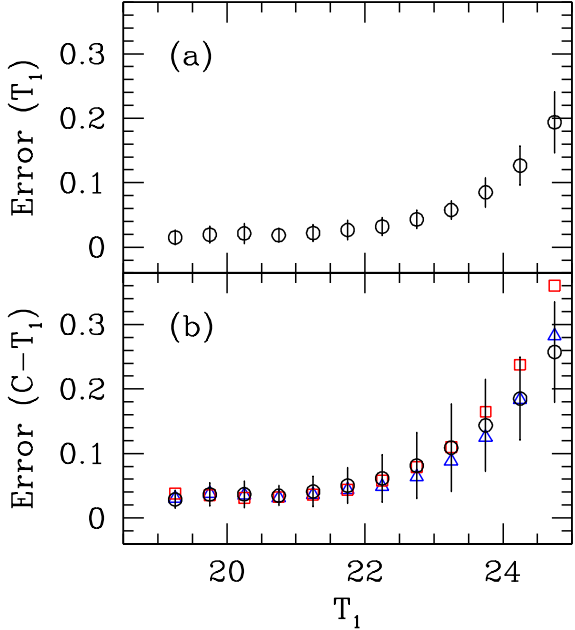}
\caption{
Mean photometric errors of $T_1$ magnitude and $(C-T_1 )$ color
for the point sources obtained from the long exposure images.
The circles with error bars, triangles, and squares represent the mean errors of all point sources, the blue GCs ($1.0<(C-T_1 ) <1.55$), and the red GCs ($1.55<(C-T_1 ) <2.1$), respectively.
\label{fig-error}}
\end{figure}

The GCs at the M86 distance appear as point sources in the KPNO images so that we derived photometry of the point sources from the images as follows. 
First, we derived isophotal model images of M86 with ELLIPSE task in IRAF/STSDAS. Then we subtracted these images from the original images to detect better point sources in the images.
Next we carried out the point-spread function (PSF) photometry on the subtracted images
  using the DAOPHOT/ALLFRAME \citep{ste94}.
Then we derived the aperture correction using the photometry of several isolated bright stars and applied it to the PSF-fitting magnitudes: 
  $0.013\pm0.006$  and $0.019\pm0.004$ for short and long exposure $C$ images
  and $-0.025\pm0.008$  and $0.022\pm0.003$  for short and long exposure $T_1$ images.
Finally, we transformed the instrumental magnitudes of the sources onto the standard system 
  using the transformation equations (April 9) given in Section 2.3. by \citet{lee08}.
We selected the point sources among the detected sources
  using the morphological classifier $r_{-2}$ moment \citep{kro80}.
We considered the sources with  $r_{-2} <1.27$ as the point sources.
Figure \ref{fig-error} displays the mean errors of $T_1$ and $(C-T_1)$ versus $T_1$ magnitude, and
Table \ref{tab-cat} lists the $CT_1$ photometry catalog of only 4351 point sources
with error $(C-T_1 )<0.3$ measured in the KPNO images.

The central region at $R\lesssim1$ arcmin in the long exposure images was saturated  and the short exposure
images are too shallow.
Therefore,
we used HST photometry available in \citet{jor09} for the analysis of the central region of M86. %
\citet{jor09} provides a photometric catalog of GCs in 100 early-type galaxies in Virgo including M86, derived from the HST/ACS Virgo Cluster Survey (ACSVCS). Their catalog includes $g$ and $z$ magnitudes, and the probability for GC classification ($p_{\rm{GC}}$). GCs appear as extended objects in the ACS images and the sources with $p_{\rm{GC}}>0.5$ 
are considered as GC candidates. 

\begin{figure}[!t]
\epsscale{0.9}
\plotone{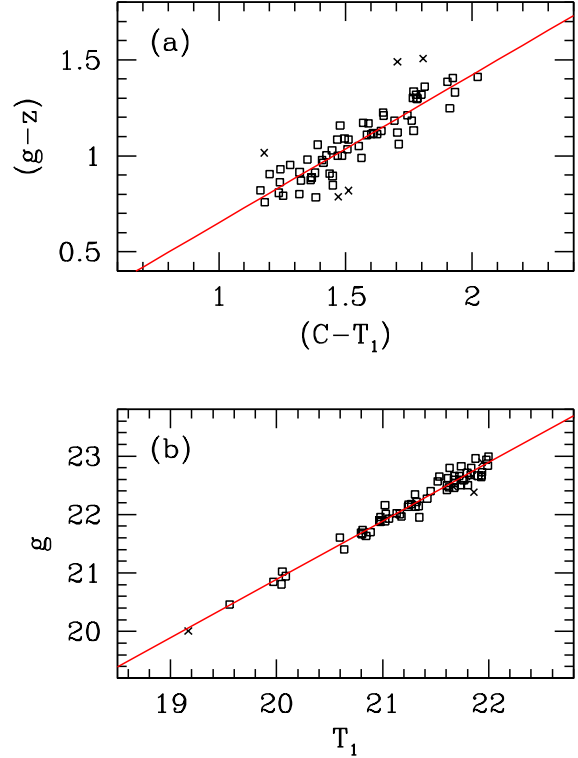}
\caption{
Comparison of $(C-T_1 )$ with $(g-z)$ (a) and $T_1$ magnitude with $g$ magnitude (b).
The solid lines represent the linear least-squares fits for error $(C-T_1 )$ $<0.08$ and $T_1<22.0$,
except for the outliers indicated by the crosses.
\label{fig-compcolmag}}
\end{figure}

We derived the transformation between ($C-T_1$) colors and ($g-z$) colors
for the bright objects (with $T_1<22.0$ mag, error $(C-T_1 ) < 0.08$ and $1.1<(C-T_1 )<2.1$), using the common objects between the KPNO results and the ACSVCS catalog.
Figure \ref{fig-compcolmag} shows relations for colors and magnitudes. 
There are several outliers, which are due to being located near  saturated sky levels or having nearby sources.
We fitted the data after excluding these outliers.
Linear fitting to the data %
yields $(g-z)=0.773 (C-T_1 ) -0.118$ with rms=0.068, and
$g=0.998 T_1 + 0.911$ with rms=0.109.

\begin{deluxetable}{ccc cc}
\tablecaption{A catalog of Washington photometry for the point sources in M86$^a$ \label{tab-cat}}
\tablewidth{0pt}
\tablehead{
\colhead{ID} &  \colhead{R.A.} & \colhead{Decl.} 
& \colhead{$T_1\pm$error} & \colhead{$(C-T_1)\pm$error} \\
\colhead{} & \colhead{(J2000)} & \colhead{(J2000)} &
\colhead{(mag)} & \colhead{(mag)} 
}
\startdata
     1 & 12:26:21.75 & 12:52:40.7 & $17.746 \pm 0.011 $&$  2.883 \pm 0.033 $ \\
     2 & 12:26:03.70 & 12:59:57.5 & $17.888 \pm 0.023 $&$  2.750 \pm 0.030 $ \\
     4 & 12:26:13.36 & 13:05:11.7 & $17.992 \pm 0.010 $&$  0.777 \pm 0.015 $ \\
     5 & 12:26:28.38 & 13:00:52.8 & $18.045 \pm 0.014 $&$  0.792 \pm 0.025 $ \\
     6 & 12:26:17.03 & 13:06:00.4 & $18.195 \pm 0.011 $&$  2.163 \pm 0.022 $ \\
    .....
\enddata
\tablenotetext{a}{The complete version of this table is in the electronic edition 
of the Journal. The printed edition contains only a sample.}
\end{deluxetable}

\subsection{Completeness of the Photometry}

\begin{figure}
\epsscale{0.9}
\plotone{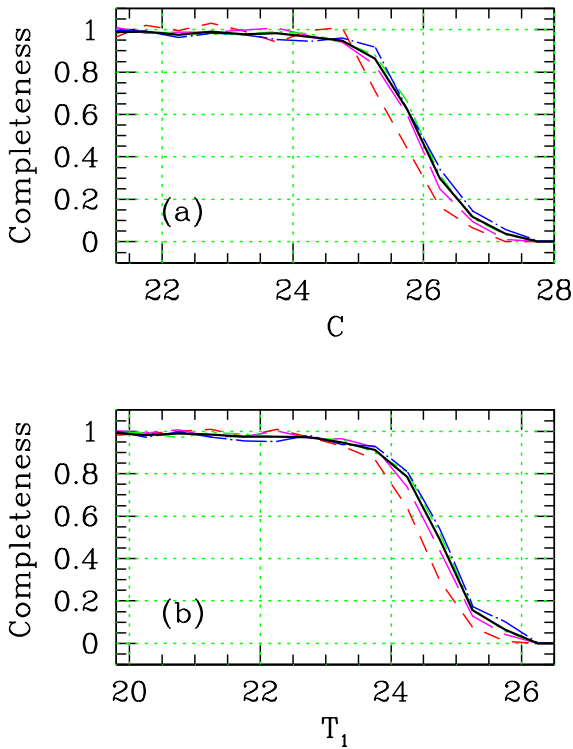}
\caption{
Completeness of $C$ magnitude (a) and $T_1$ magnitude (b).
The solid lines represent the results for the entire region. 
The short dashed, long dashed, dot-short dashed, and dot-long dashed lines indicate
the completeness at $1\arcmin<R<3\arcmin$, $3\arcmin<R<5\arcmin$, 
$5\arcmin<R<7\arcmin$, and $7\arcmin<R<9\arcmin$, respectively.
\label{fig-complete}}
\end{figure}

We estimated the completeness of the KPNO photometry
  using DAOPHOT/ADDSTAR, which is designed for the artificial star experiment.
First we generated a set of artificial stars of which
   the color-magnitude diagrams and luminosity functions are similar to those of the detected point sources,  
   using the PSFs derived from the long exposure  images.
Next we added them to the real image avoiding the position of detected real sources.
  We obtained 50 artificial images each of which includes 1200 artificial stars for each filter
  so that the total number of added artificial stars is 60,000.
Then we applied the same photometry procedure to the artificial images as used for the real image.
From these results,
we estimated the completeness, which is the number ratio of the recovered artificial stars
to the added artificial stars.

Figure \ref{fig-complete} displays the completeness of $C$ and $T_1$ magnitude measured for the point sources.
This figure shows the completeness for the entire region and four radial bins:
$1\arcmin<R<3\arcmin$, $3\arcmin<R<5\arcmin$,  $5\arcmin<R<7\arcmin$, and $7\arcmin<R<9\arcmin$.
The 90 \% completeness is located at %
   $T_1=23.8$ ($C=25.1$) for the outer region at $R>3\arcmin$ 
  and $T_1=23.5$ ($C=24.9$) for the inner region at $1\arcmin<R<3\arcmin$.
The photometric limit levels with 50 \% completeness are 
  $T_1=24.7$ ($C=25.9$) for the outer region at $R>3\arcmin$ and
  $T_1=24.4$ ($C=25.6$) for the inner region at $1\arcmin<R<3\arcmin$.
The completeness varies little, depending on galactocentric radius for the outer region at $R>3\arcmin$,
while it is somewhat lower for the inner region at $R<3\arcmin$ compared with the outer region.

\section{RESULTS}

\subsection{Color-Magnitude Diagram}

\begin{figure*}[t]
\centering
\epsfxsize=14cm
\epsfbox{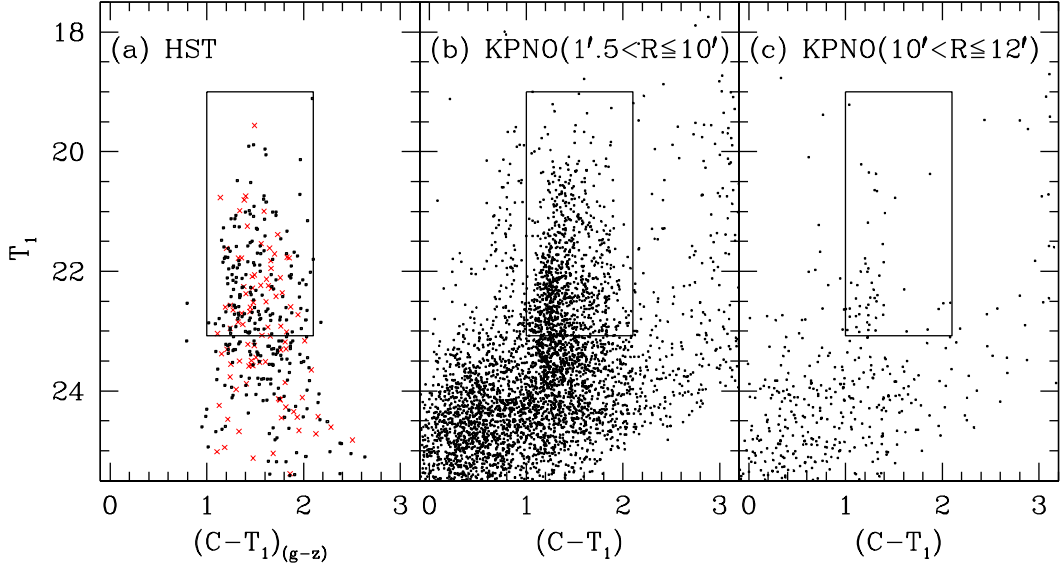}
\caption{
Color-magnitude diagrams of 
the sources in M86 given in ACSVCS catalog (a)
and the point sources in M86 %
measured from the KPNO images (b and c).
The boxes indicate the color ($1.0<(C-T_1 )<2.1$) and magnitude ($19<T_1<23.08$) criteria for selecting the M86 GCs. 
The dotted and crosses in panel (a) represent the values
converted from $g$ and $(g-z)$ at $R<1\arcmin.5$ and $1\arcmin.5<R<2\arcmin.4$, respectively.
\label{fig-cmd}}
\end{figure*}

Figure \ref{fig-cmd} displays the color-magnitude diagrams (CMDs) of the point sources 
measured from the KPNO images of M86
 as well as that from the HST/ACS photometry (for $R<1\arcmin.5$).
 We plotted the CMDs for the M86 region ($1\arcmin.5<R<10\arcmin$) and the background region ($10\arcmin < R < 12\arcmin$).
 For the central region at $R<1\arcmin.5$ we plotted only the HST photometry.   
We transformed the ($g-z$) colors and $g$ magnitudes of the objects given in the ACSVCS catalog 
  into ($C-T_1$) colors and $T_1$ magnitudes, respectively, 
  using the transformation equations derived in the previous section.

Three distinct features are seen in Figure \ref{fig-cmd}:
(a) GCs of M86  %
  in the broad vertical feature in the color range of $1.0<(C-T_1)<2.1$,
(b) a small number of bright foreground stars
at $(C-T_1) \lesssim 1.0$ and $(C-T_1) \gtrsim 2.5$,
and
(c) faint blue unresolved background galaxies with $T_1>23$ mag.
We selected the bright point sources with $1.0<(C-T_1)<2.1$  and $19<T_1 <23.08$ 
  for the analysis of the M86 GCs,
considering the following points:
(1) the color range of the known GCs in other galaxies is $(C-T_1 ) \approx 0.9$ to 2.1 \citep{gei96,lee98,dir05, for07,lee08};
(2) the peak luminosity of the M86 GCs is $T_1= 23.08$,
  which was converted from the peak $g$ magnitude ($\mu_g=23.95\pm0.10$) given by \citet{jor07}; 
(3) the mean photometric error and the completeness for $T_1 <23.08$ are 
  smaller than error $(T_1 ) = 0.05$ and larger than 90\%, respectively; and
(4) the contamination due to background galaxies is estimated to be very small for $T_1 <23.08$.

\subsection{Color Distribution of the Globular Clusters}

\begin{figure}[!t]
\epsscale{1.0}
\plotone{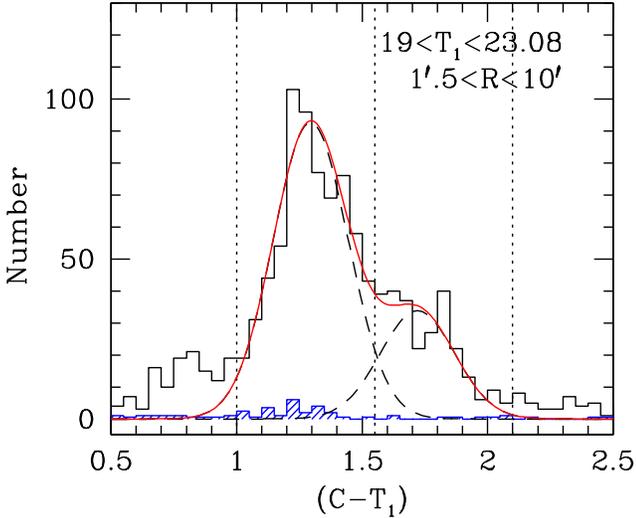}
\caption{
Color distribution of the M86 GCs with $19<T_1<23.08$ and $1\arcmin.5<R<10\arcmin$.
The solid line and dashed lines represent the double Gaussian fit derived from the KMM test.
The hashed histogram is the color distribution of the point sources  within the background region ($10\arcmin<R<12\arcmin$).
The vertical dotted lines indicate the color boundaries for the blue GCs ($1.0<(C-T_1 ) <1.55$) and the red GCs ($1.55<(C-T_1 ) <2.1$).
\label{fig-coldist}}
\end{figure}

\begin{figure}[!t]
\plotone{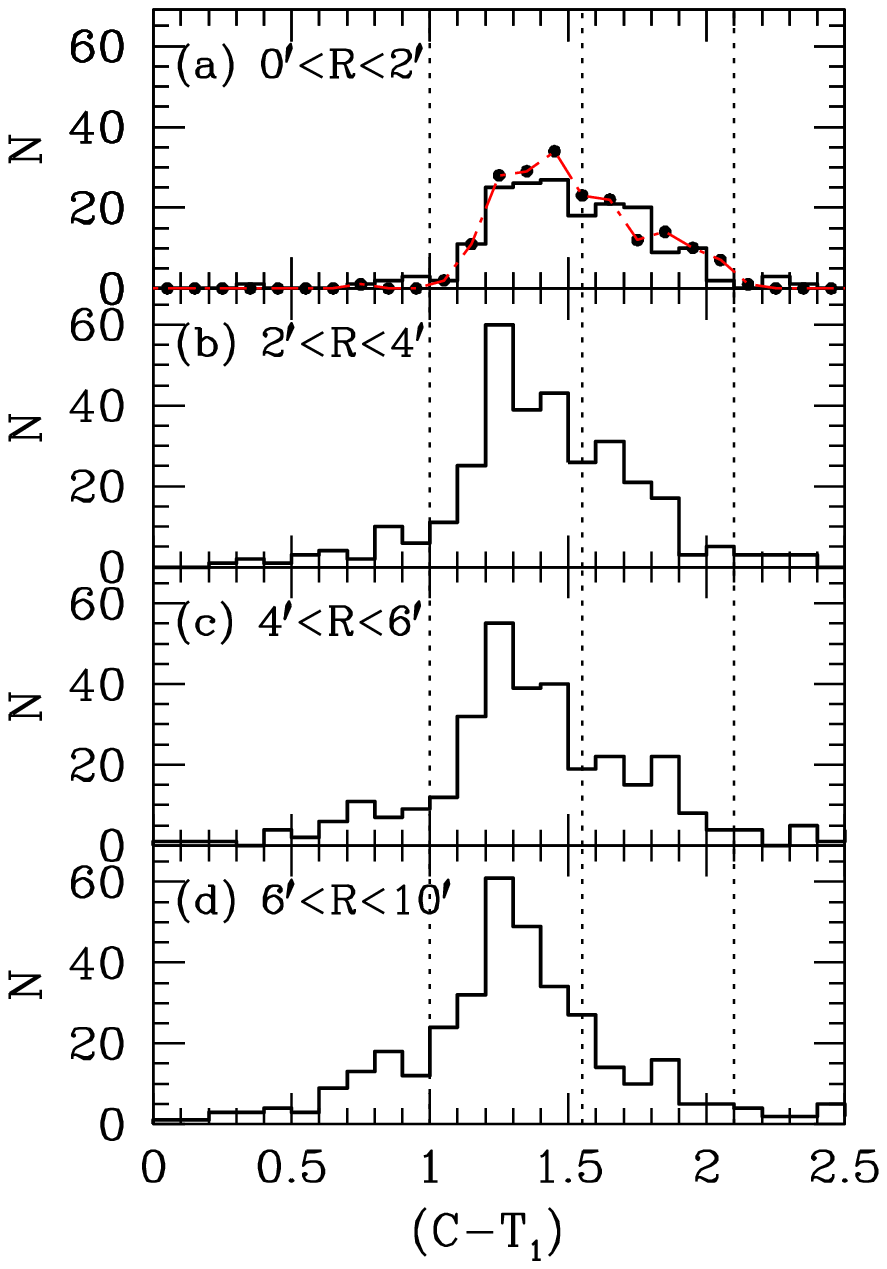}
\caption{
Radial color distribution of the GCs in M86.
The histograms in each panel and the dot-dashed line with the circles in panel (a) 
represent the color distributions of the GCs obtained from KPNO and from ACSVCS catalog, respectively.
The vertical dotted lines indicate the color boundaries for the sub-populations.
\label{fig-radcol}}
\end{figure}

Figure \ref{fig-coldist} shows the color distribution of the point sources
with $19<T_1 <23.08$ at $1\arcmin.5 <R< 10 \arcmin$ in the KPNO images.
The color distribution of the background objects with the same
magnitude range at $10\arcmin<R<12\arcmin$ is also shown.

We carried out the KMM test, 
  which estimates the statistical significance of bimodality in astronomical datasets \citep{ash94},
  to investigate the bimodality of the $(C-T_1 )$ color distribution of the M86 GCs.
KMM test of the data shows that the probability for bimodal color distribution is higher than 99.9 \%.
The color distribution is also well described as the best-fit double Gaussian curves:
a blue component with center at $(C-T_1 )=1.30$ and width $\sigma=0.15$ and 
a red component with center at $(C-T_1 )=1.72$ and width $\sigma=0.15$. 
The minimum between the two components is found to be at $(C-T_1)\approx1.55$.
Thus we divided the entire sample of the M86 GCs into two sub-populations:
the blue GCs (BGC) with $1.0<(C-T_1)<1.55$ and the red GCs (RGC) with $1.55<(C-T_1)<2.1$.
This boundary color is consistent with the previous study of M86 \citep{rho04},
but is bluer than that used for M60 \citep{lee08} and M49 \citep{lee98}
and slightly redder than that used for M87 \citep{cot01}.
Figure  \ref{fig-radcol} displays the radial variation of the color distribution of the M86 GCs.
The number ratio of the blue GCs and the red GCs
increases as the galactocentric radius increases.

\subsection{Surface Photometry of M86}

We derived surface photometry of M86 from the KPNO images using the IRAF/ELLIPSE task. 
After masking out the bright foreground stars and background galaxies and
  using the mode values of the regions at $R \gtrsim 10\arcmin$ in the north-east corner as the background levels,
  we carried out  ellipse fitting to the $C$ and $T_1$ images.
The errors for the surface brightness magnitudes %
are given by ELLIPSE.
The long exposure images are saturated at the central region of M86,
  while the short exposure images are not.
Thus we used the values from the short exposure images at $R< 1 \arcmin$, 
  the average values from the short and long exposure images at $1\arcmin <R< 3 \arcmin$, and
  the values from the long exposure images at $R> 3\arcmin$.

\begin{figure*}[t]
\centering
\epsfxsize=14cm
\epsfbox{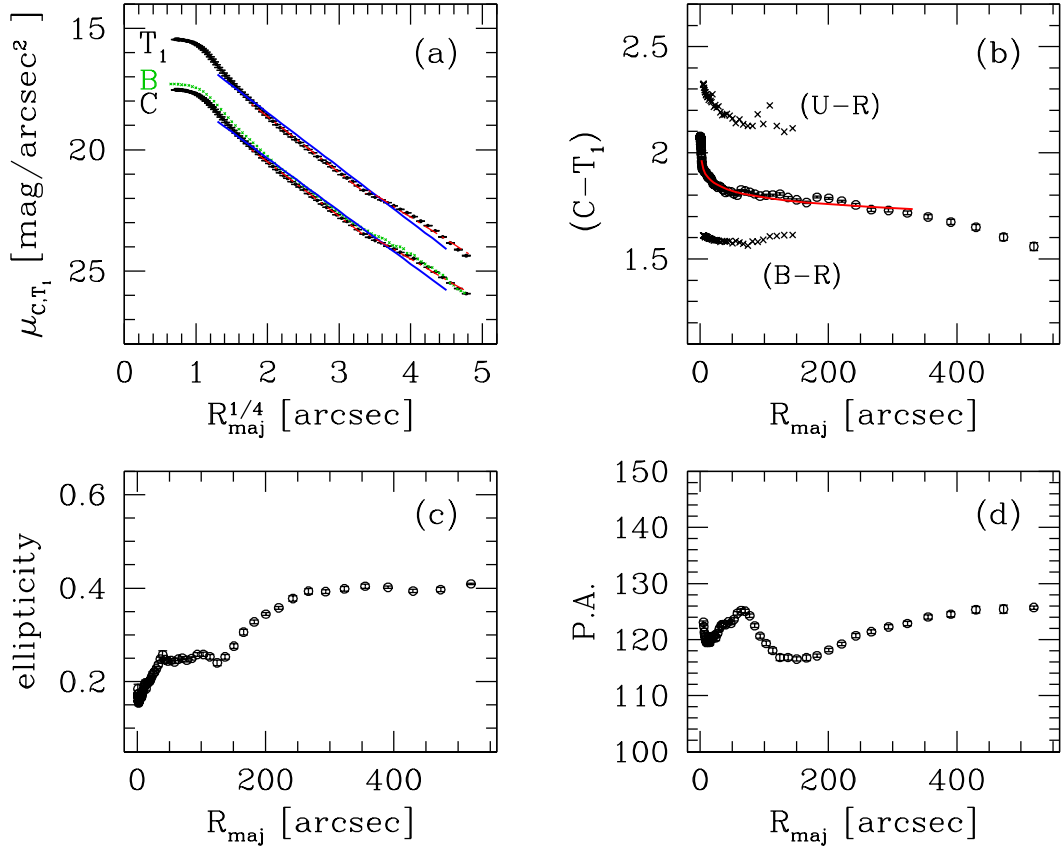}
\caption{
Surface photometry of M86.
$R_{maj}$ is galactocentric distance along the major axis of M86 stellar light.
(a) Surface brightness magnitudes for $T_1$ and $C$.
The solid lines represent the linear least-squares fits with $R^{1/4}$ law for  $2\arcsec<R_{maj}<400\arcsec$. 
The dot-dashed lines represent the double linear fits with $R^{1/4}$ law 
  for $10\arcsec<R_{maj}<135\arcsec$ and $135\arcsec<R_{maj}<560\arcsec$.
The crosses indicate the surface brightness profile for $B$ magnitude given by \citet{cao90}.
(b) Surface color profiles. 
The solid line indicates the power law fit of $(C-T_1 )$ and log$(R)$ for $2\arcsec<R_{maj}<300\arcsec$.
The circles and crosses represent the colors from this study and from \citet{pel90}, respectively.
(c) Ellipticity versus $R_{maj}$.  (d) Position angle (P.A.)  versus $R_{maj}$.
\label{fig-surf}}
\end{figure*}

Figure \ref{fig-surf} shows the radial profiles of the surface brightness magnitudes, $(C-T_1)$ color,
 ellipticity, and position angle (P.A.) of M86 as a function of major radius $R_{maj}$.
The surface brightness profiles of the $C$ and $T_1$ magnitudes are consistent with
  that of the $B$ magnitude given by \citet{cao90}.
The surface brightness profiles are roughly described by a de Vaucouleurs  law.
These surface brightness profiles are fitted with a de Vaucouleurs  $R^{1/4}$ law 
 for the range of $2\arcsec <R_{maj}<400\arcsec$, using  linear least-squares fitting:
$\mu(C)= 2.173(\pm 0.032) R^{1/4} + 16.009(\pm 0.086)$ with rms=0.219,
and $\mu(T_1 )= 2.247(\pm 0.033) R^{1/4} + 13.981(\pm 0.089)$ with rms=0.226.
The effective radius ($R_{eff}$) and standard radius ($R_{25}$) for $C$-band 
are derived to be $3\arcmin.59$ and $4\arcmin.88$ 
(3$\arcmin.14$ and 9$\arcmin.64$ for $T_1$),
corresponding to linear sizes of 17.54 kpc and 23.85 kpc,  respectively.

This value for the effective radius is %
  much smaller than that derived from the small field HST/ACS images by \citet{fer06} 
  ($R_{\rm eff}=6\arcmin.86 \simeq 33.52$  kpc at $g$-band)
 but similar to that derived from wide field images by \citet{kor09} ($R_{\rm eff}=3\arcmin.38 \simeq 16.52$  kpc at $V$-band). 
The surface brightness profiles of $C$ and $T_1$ magnitudes in this study as well as  the $B$ magnitude given by \citet{cao90}
show a break at $R_{maj} \sim 135\arcsec$, resulting in flatter profiles in the outer region. 
Thus we carried out double linear fits with $R^{1/4}$ law 
  at $10\arcsec<R_{maj}<135\arcsec$: 
    $\mu(C)   = 2.240(\pm 0.016) R^{1/4} + 16.027(\pm 0.042)$ with rms=0.038
and $\mu(T_1 )= 2.304(\pm 0.021) R^{1/4} + 14.030(\pm 0.052)$ with rms=0.047,
   and $135\arcsec<R_{maj}<560\arcsec$:
    $\mu(C)   = 1.706(\pm 0.042) R^{1/4} + 17.684(\pm 0.170)$ with rms=0.063
and $\mu(T_1 )= 1.860(\pm 0.056) R^{1/4} + 15.338(\pm 0.227)$ with rms=0.084.  

The $(C-T_1)$ color of the M86 nucleus is very red with 
$(C-T_1) \approx 2.1$. It becomes rapidly bluer in the inner region ($R_{maj}\leq50\arcsec$) 
as the galactocentric radius increases, 
but changes slowly in the outer region. 
This trend %
is consistent with the radial variation  of
$(U-R)$ colors given by \citet{pel90} as shown in Figure \ref{fig-surf} (b),
but is somewhat different from the $(B-R)$ colors given by \citet{pel90}. 
This may be related with the existence of dust close to the M86 center, as seen in the far-infrared images \citep{sti03,gom10}.
The color at $2\arcsec<R_{maj}<300\arcsec$ is approximately fitted by the log-linear
relation $(C-T_1 )=-0.105(\pm 0.003) {\rm \log} R + 1.999(\pm0.005)$ with rms=0.014.
The ellipticity increases rapidly from 0.15  in the central region to 0.25 at $R_{maj} \approx 40\arcsec$, staying almost constant
at $40\arcsec<R_{maj}< 135\arcsec$. Then it increases again to $\sim 0.4$ in the outer region. 
Thus the isophotes of the outer region
are more elongated than the central region. 
The values of the ellipticity at the effective radius and standard radius
are 0.35 and 0.40, respectively. 
The value for P.A. is constant with $120\pm6$ deg.

\subsection{Spatial Distribution of the Globular Clusters}

\begin{figure*}[t]
\centering
\epsfxsize=14cm
\epsfbox{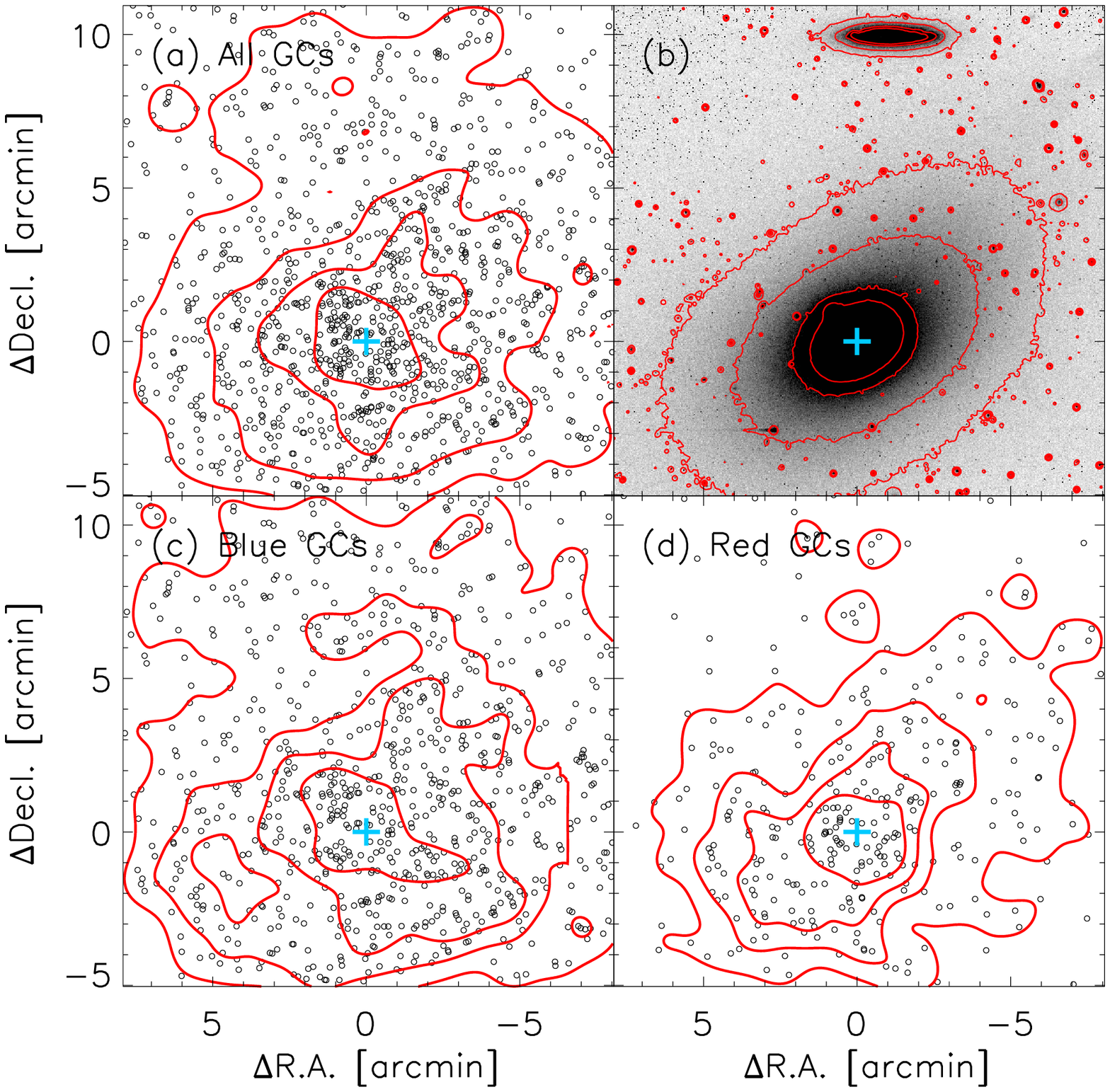}
\caption{
Spatial distributions and number density contours of M86 GCs. 
The plus signs indicate the center of the galaxy.
(a) All GCs. The contour levels are 2.037,  5.093,  8.148, and 12.223 GCs arcmin$^{-2}$.
(b) The grey map with isophotal contours in the $T_1$ image of M86.
(c) Blue GCs. The contour levels are 0.6 times of those for all the GCs.
(d) Red GCs. The contour levels are 0.4 times of those for all the GCs.
\label{fig-spatcon}}
\end{figure*}

We investigated the spatial structure of the M86 GC system.
Figure \ref{fig-spatcon} displays the spatial distributions of the GCs
with $19<T_1<23.08$ mag (all GCs, blue GCs, and red GCs)
as well as M86 stellar light.
We also overlay number density contours of the GCs in this figure.
$T_1$ image of M86 with isophotal contours is displayed in the panel (b) for comparison with the GC system.

Several features are noted in this figure. %
First, all the GCs have a circular spatial distribution,  
  and show a central concentration.
Second, the blue GCs have a roughly circular spatial distribution,  extended farther than the red GCs.
Third, the red GCs have a spatial distribution somewhat elongated along the southeast-northwest direction,
  which is consistent with the distribution of M86 stellar halo.
Using the dispersion ellipse method \citep{tru53, hwa08, str11},
we measured $\epsilon$ (ellipticity)= $0.04\pm0.02$, $0.02\pm0.03$, and $0.16\pm0.05$, and
 P.A.=$122\pm24$, $42\pm43$, and $131\pm10$ deg
 for all the GCs, the blue GCs, and the red GCs at $R< 5\arcmin$, respectively.
 The value for the ellipticity of the red GCs is eight times larger than that of the blue GCs. 
The P.A.'s of all the GCs and the red GCs are similar to that of M86 stellar light.

\subsection{Radial Number Density Profiles of the Globular Clusters}

\begin{figure}[!t]
\plotone{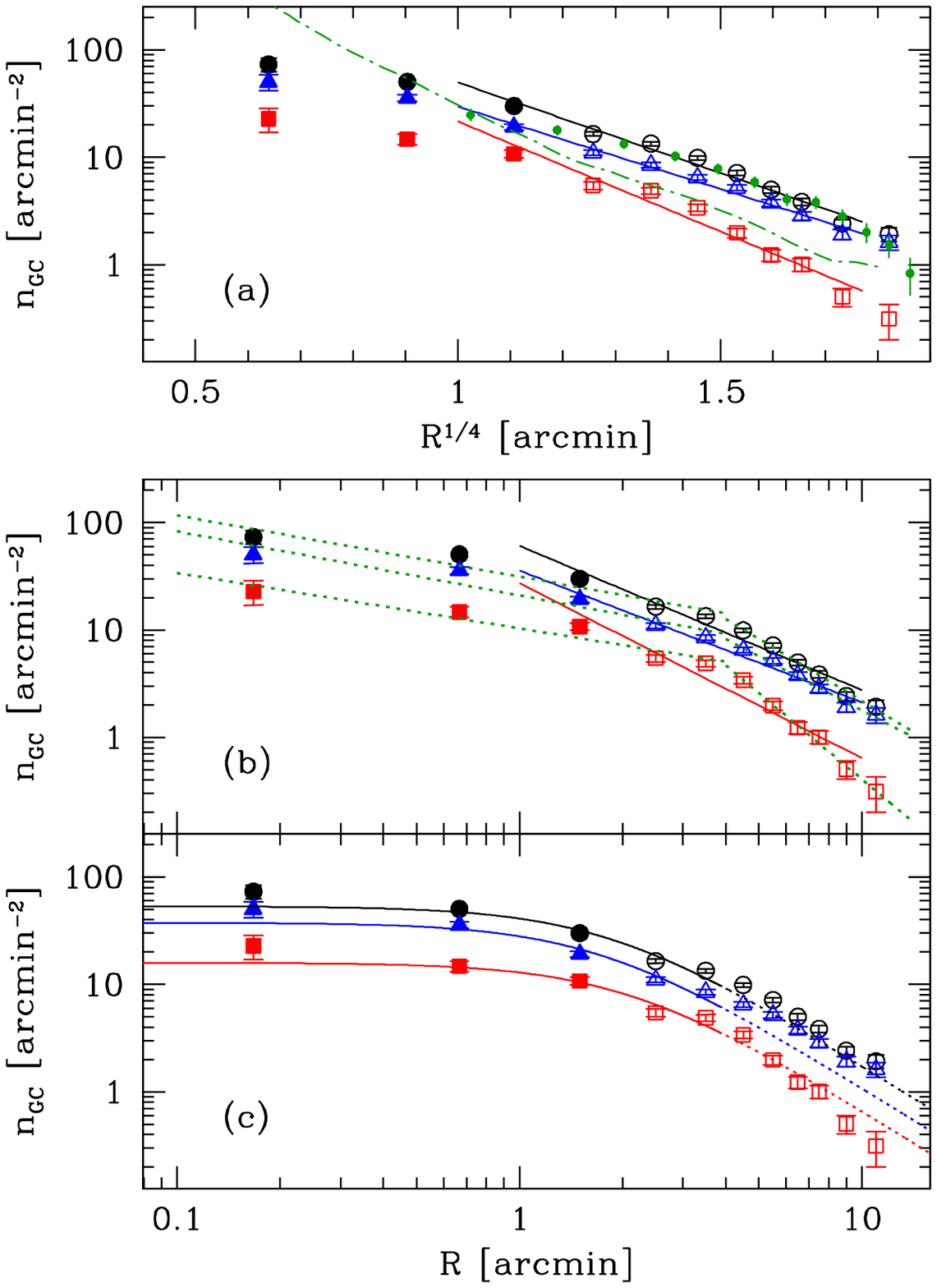}
\caption{
Radial number density profiles of M86 GCs.
The filled and open points are from HST and KPNO, respectively.
The circles, triangles, and squares represent all the GCs, the blue GCs, and the red GCs, respectively.
(a) Number density versus $R^{1/4}$.
  The small filled circles indicate the number density profile given by \citet{rho04}.
  The solid lines represent the fits for a de Vaucouleurs law for $1\arcmin<R<10\arcmin$.
  The dot-dashed line indicates the $C$ surface brightness profile of the M86 stellar halo.
(b) Number density versus $R$.
  The solid lines represent the fits for a power law.
  The dotted lines indicate the two-component power law fits for  $0\arcmin.1<R<4\arcmin$ and $4\arcmin<R<12\arcmin$.
(c) Number density versus $R$.
The solid lines represent King model fits for $R<4\arcmin$, %
which are represented by the dotted lines for $R>4\arcmin$.
\label{fig-rden}}
\end{figure}

We derived the radial number density profiles of the GCs with $19<T_1<23.08$ mag, 
  using the KPNO results for the outer region at $R>1\arcmin.5$ and
  the ACSVCS catalog for the central region at $R<1\arcmin$.5.
Here we masked out the region with $5\arcmin\times 1\arcmin.5$
centered on NGC 4402 to remove any contamination due to  this galaxy.
First, we applied the correction for the incompleteness of photometry using the results shown in Figure \ref{fig-complete}.
\citet{rho04} found that the radial number density of the GCs in M86 becomes zero at
$R>17\arcmin$. The field of view of our KPNO images is not wide enough to cover the entire region of M86.   Here we adopted, as a background number density, %
a half of the number density of the point sources at $10\arcmin<R<12\arcmin$:
$0.915\pm 0.279$ arcmin$^{-2}$ for all  the GCs, 
$0.767\pm 0.255$ arcmin$^{-2}$ for the blue GCs, and 
$0.149\pm 0.113$ arcmin$^{-2}$ for the red GCs. 
We subtracted these values from the original counts. 
Then we derived radial number densities, considering  the effect of the GC luminosity function (by multiplying by a factor of two).
The radial number densities of the GCs thus derived are listed in Table \ref{tab-rden}.

Figure \ref{fig-rden} displays the radial number density profiles of the GCs versus $R^{1/4}$ in (a) and $R$ in (b) and (c).
We also plot the radial number density profile of M86 GCs given by \citet{rho04} and the $C$-band surface brightness profile of M86 in (a) for comparison.
Several notable features are seen as follows. %
First, the density profile of all the GCs is more extended than the galaxy stellar light. 
  The density profile of all the GCs  derived in this study agrees well with that given by \citet{rho04}. 
Second, the density profile of the blue GCs decreases more slowly as the galactocentric distance increases 
  than that  of the red GCs.
Third, the density profile of the red GCs is consistent with 
  the surface brightness profile of the galaxy stellar light. 

Fourth, the density profiles of the GCs at $1\arcmin<R<10\arcmin$ are approximately fitted  %
 either by a de Vaucouleurs  $R^{1/4}$ law:
$\log n_{GC} = -1.683(\pm0.104) R^{1/4} + 3.379(\pm 0.154) $ for all the GCs, 
$\log n_{GC} = -1.534(\pm0.082)$ $R^{1/4} + 3.006(\pm 0.120) $ for the blue GCs, and
$\log n_{GC} = -2.048(\pm0.182) R^{1/4} + 3.382(\pm 0.268) $ for the red GCs;
 or by a power law:
$\log n_{GC} = -1.343(\pm0.116)  \log R + 1.783(\pm 0.080) $ for all the GCs,
$\log n_{GC} = -1.226(\pm0.094)$ $\log R + 1.552(\pm 0.064) $ for the blue GCs, and 
$\log n_{GC} = -1.628(\pm0.185)  \log R + 1.434(\pm 0.127) $ for the red GCs.
In the case of all the GCs, the values for the slope agree well with those given by \citet{rho04}.
Effective radii of all the GCs, the blue GCs, and the red GCs are derived from the de Vaucouleurs  $R^{1/4}$ law fits:
$15\arcmin.33$ ($= 74.50$ kpc), $22\arcmin.20$ ($= 107.89$ kpc), and $ 6\arcmin.99$ ($= 33.97$ kpc), respectively. Thus the effective radius of the red GCs is much smaller than that
of the blue GCs, showing the central concentration of the red GCs is much stronger than that of the blue GCs.
The density profiles of the GCs show
a break at  $R \approx 4\arcmin$, resulting steeper profiles
in the outer region.
The outer parts at $R>4\arcmin$ are fitted well by a power law:
$\log n_{GC} = -1.915(\pm0.085)  \log R + 2.255(\pm 0.073) $ for all the GCs,
$\log n_{GC} = -1.672(\pm0.101)  \log R + 1.923(\pm 0.086) $ for the blue GCs, and 
$\log n_{GC} = -2.666(\pm0.109)  \log R + 2.276(\pm 0.093) $ for the red GCs.
The inner parts at $R<4\arcmin$ are fitted well by a power law:
$\log n_{GC} = -0.569(\pm0.094)  \log R + 1.496(\pm 0.045) $ for all the GCs,
$\log n_{GC} = -0.596(\pm0.100)  \log R + 1.322(\pm 0.047) $ for the blue GCs, and 
$\log n_{GC} = -0.516(\pm0.089)  \log R + 1.014(\pm 0.042) $ for the red GCs.
In the case of the outer parts, 
 the radial number density profile of the red GCs is much steeper than that of the blue GCs,
 while
 in the case of the inner parts,
 the sub-populations have similar slopes.

Finally, the density profiles of the GCs are somewhat flat in the central region at $R\lesssim 2\arcmin$. 
This flat radial profile, which may indicate dynamical relaxation of the GC system \citep{cot98} or the relic of the GC formation epoch \citep{har98,kun99}, 
can be approximately fitted by a King model \citep{kin62}.
Figure \ref{fig-rden} (c) displays the results of a King model fitting for $R<4\arcmin$.
We derived core radius
$r_c=1\arcmin.82$ and concentration parameter $c=5.13$ for all the GCs,
$r_c=1\arcmin.71$ and $c=5.34$ for the blue GCs, and
$r_c=2\arcmin.07$ and $c=5.68$ for the red GCs.
It is noted that in the case of the blue GCs
  the radial density profile of the outer region at $R>4\arcmin$ shows an excess over the King model.

\begin{deluxetable}{cccc} 
\tablecaption{Radial number density profiles of the GCs in M86\label{tab-rden}}
\tablewidth{0pt}
\tablehead{
\colhead{$R$} & \colhead{$n_{\rm{AGC}}$} & \colhead{$n_{\rm{BGC}}$} & \colhead{$n_{\rm{RGC}}$} \\
\colhead{(arcmin)} & \colhead{(arcmin$^{-2}$)}& \colhead{(arcmin$^{-2}$)} & \colhead{(arcmin$^{-2}$)}
}
\startdata
  0.167 & $  73.193\pm 10.417$ & $  50.392\pm  8.668$ & $  22.800\pm  5.778$ \\
  0.667 & $  50.465\pm  3.066$ & $  35.706\pm  2.588$ & $  14.758\pm  1.644$ \\
  1.500 & $  29.959\pm  1.487$ & $  19.182\pm  1.201$ & $  10.776\pm  0.877$ \\
  2.500 & $  16.475\pm  0.755$ & $  11.020\pm  0.626$ & $   5.455\pm  0.423$ \\
  3.500 & $  13.403\pm  0.589$ & $   8.506\pm  0.478$ & $   4.897\pm  0.343$ \\
  4.500 & $   9.894\pm  0.454$ & $   6.498\pm  0.376$ & $   3.394\pm  0.254$ \\
  5.500 & $   7.169\pm  0.383$ & $   5.191\pm  0.332$ & $   1.979\pm  0.192$ \\
  6.500 & $   4.993\pm  0.324$ & $   3.761\pm  0.285$ & $   1.232\pm  0.153$ \\
  7.500 & $   3.854\pm  0.286$ & $   2.849\pm  0.252$ & $   1.006\pm  0.137$ \\
  9.000 & $   2.422\pm  0.224$ & $   1.917\pm  0.202$ & $   0.505\pm  0.097$ \\
 11.000 & $   1.927\pm  0.279$ & $   1.614\pm  0.255$ & $   0.314\pm  0.113$ \\
\enddata
\end{deluxetable}

\subsection{Radial Variation of Globular Cluster Colors} 

\begin{figure}[!t]
\plotone{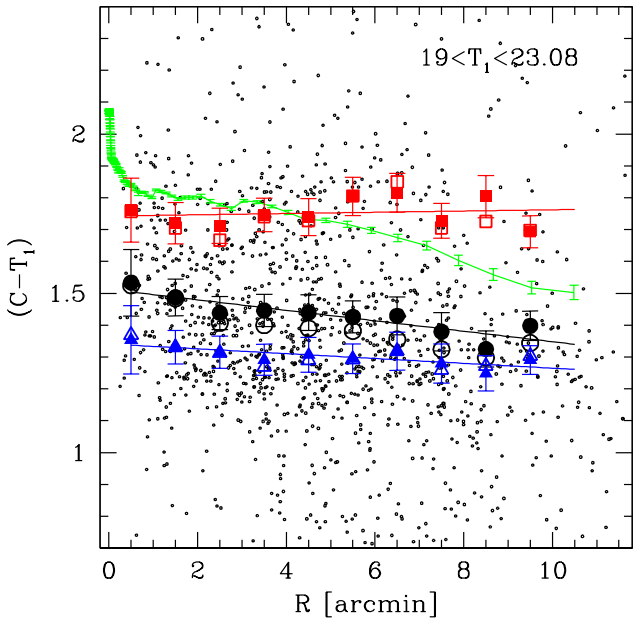}
\caption{
$(C-T_1 )$ color of M86 GCs as a function of galactocentric distance.
The circles, triangles, and squares represent all the GCs, the blue GCs, and the red GCs, respectively.
The filled and open points represent mean and median color of the GCs in each radial bin, respectively.
The solid lines are the linear least-squares fits about the mean color of the sub-populations.
The curved line indicates $(C-T_1 )$ surface color profile of M86 stellar halo.
\label{fig-rcol}}
\end{figure}

We investigated the radial variation of the color of the M86 GCs.
Figure \ref{fig-rcol} displays the $(C-T_1)$ colors of the GCs with $19<T_1<23.08$ mag
as a function of galactocentric distance derived from the KPNO images.
The mean color of all the GCs in each radial bin shows a radial gradient, 
  while those of the blue GC and the red GC show little radial gradients.
The linear least-squares fitting at  $R< 10\arcmin$ yields
$(C-T_1)=-0.016(\pm 0.003) R + 1.512(\pm 0.018)$ with rms=0.026 for all the GCs,
$(C-T_1)=-0.008(\pm 0.002) R + 1.341(\pm 0.012)$ with rms=0.017 for the blue GCs, and
$(C-T_1)= 0.002(\pm 0.005) R + 1.743(\pm 0.028)$ with rms=0.040 for the red GCs.
The color gradient of all the GCs is due to the variation of the number ratio of
  the blue GCs to the red GCs as shown in Figure  \ref{fig-radcol}.
  This negative color gradient for all the GCs is consistent with the previous studies based on $(B-R)$ color \citep{rho04}. %
The mean color of the M86 stellar light is much closer to that of the red GCs than to that of the blue GCs, as often seen
in other gEs such as M49 and M60  \citep{lee98, lee08}.
%

\section{DISCUSSION} 

\begin{figure}[!t]
\plotone{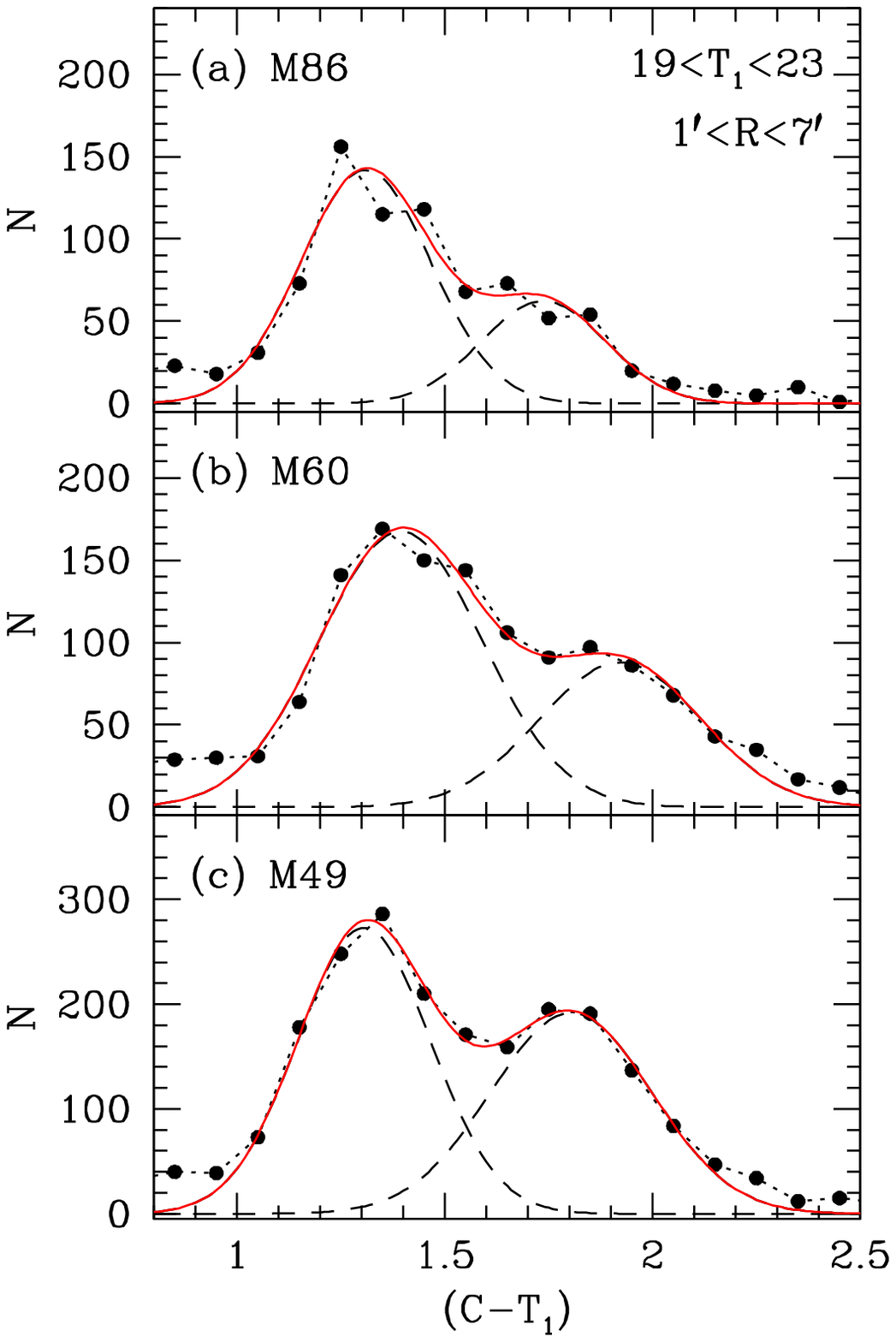}
\caption{
Comparison of the color distribution between the GCs in M86 and the GCs in other Virgo gEs.
(a) M86 from this study. (b) M60 from \citet{lee08}. (c) M49 from \citet{gei96} and \citet{lee98}.
The solid and dashed lines represent the double Gaussian fits.
\label{fig-gE3col}}
\end{figure}

We compared the photometric results of the M86 GC system  with those of GC systems in other two gEs in Virgo (M60 and M49).
  M60, which is the third brightest gE in Virgo and  located about 3 deg from the  Virgo center, and 
  M49, which is the brightest gE in Virgo and located about 4 deg from the  Virgo center.
 Previous studies for the GC systems of M60 \citep{lee08} and M49 \citep{gei96b,lee98}
 used the same instruments and similar photometric techniques as this study for the M86 GC system.
For comparison we used the same criteria for the GC sample in three galaxies, GCs with $R<7\arcmin$ and $19<T_1<23$.

Figure \ref{fig-gE3col} displays the color distributions of the GCs in M86, M60, and M49. 
The GCs in these galaxies show similarly bimodal color distributions, but 
 there are slight differences among these galaxies.
The peak colors for M86 ($(C-T_1)=1.31$ and 1.74) are similar to those for M49 ($(C-T_1)=1.30$ and 1.79),
  but are about 0.1 mag bluer than those for M60 ($(C-T_1)=1.37$ and 1.87).
Applying this color difference to the relation between $(C-T_1)$ color and metallicity \citep{lee08},   
  this implies that the M86 and the M49 GCs are on average about 0.2 dex more metal-poor than M60 GCs.
These results are approximately consistent with those by \citet{pen06} who covered the central regions with $R\lesssim 1\arcmin.5$.
They reported that the peak colors for these three galaxies are
  $(g-z)=0.98$ and 1.33 for M86, $(g-z)=0.97$ and 1.42 for M49, and $(g-z)=0.98$ and 1.45 for M60.

There are significant differences among the number ratios (N(BGC)/N(RGC)) of the blue GCs to the red GCs in these galaxies:
  $2.37\pm 0.26$ for M86, $1.38\pm 0.11$ for M60, and $0.97\pm 0.07$ for M49.
Thus the fraction of the blue GCs for M86, the most faint among these three galaxies, is the largest.
This is consistent with previous finding that the fraction of the blue GCs decreases as their host galaxy gets brighter \citep{pen06}.
On the other hand,
the number ratio of the blue GCs to the red GCs in each galaxy is a little bit larger than the value by \citet{pen06} (
$2.33\pm 1.15$ for M86, $0.75\pm 0.10$ for M60, and $0.69\pm 0.09$ for M49) and
by \citet{fai11} ($1.07\pm 0.10$ for M60).
The radial coverages used by \citet{pen06} and \citet{fai11} are $R\lesssim 1\arcmin.5$ and
  $R\lesssim 5\arcmin$, respectively, so that their fields are on average closer to the galaxy center
  than that used in this study ($1\arcmin<R<7\arcmin$). 
These shows that   %
 this number ratio decreases as the galactocentric radius decreases (see Section 3.2 for detail).

\begin{figure*}[t]
\centering
\epsfxsize=14cm
\epsfbox{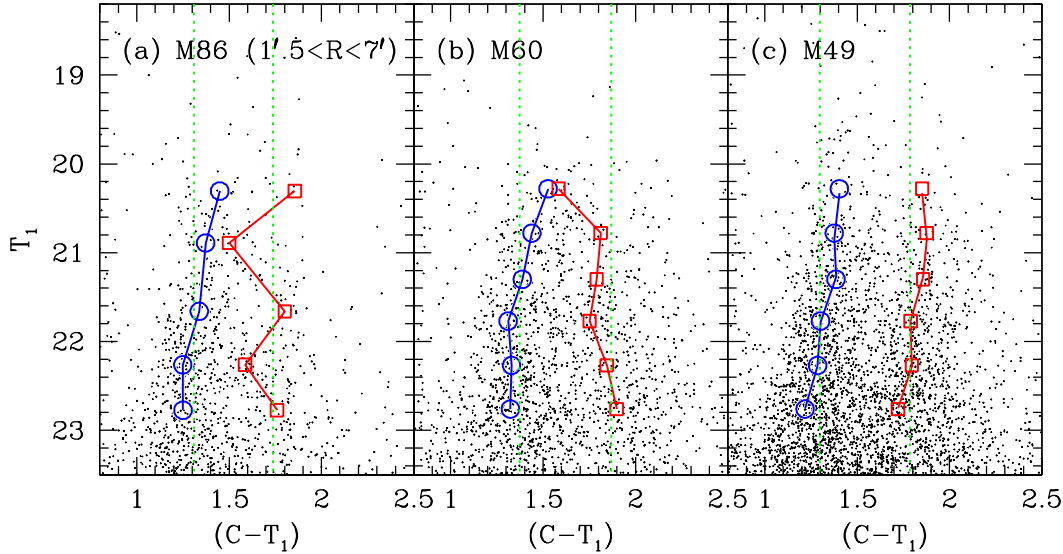}
\caption{
Comparison of the CMDs for the GCs in the Virgo gEs with $1\arcmin.5<R<7\arcmin$.
(a) M86. (b) M60. (c) M49.
The vertical dotted lines indicate the peak values of the blue GCs and the red GCs with $19<T_1<23$. %
The large open circles and squares represent the peak values in each magnitude bin derived
 from the KMM test.
\label{fig-gE3cmd}}
\end{figure*}

Recent studies for the GCs in gEs 
  based on HST/ACS data  \citep{har06,str06, mie06, pen09, mie10} and
  wide-field images \citep{for07, lee08, har09},
  found a so-called `blue tilt' (or a color-magnitude relation):
  the brighter the bright blue GCs are, the redder their colors get.
   Several mechanisms were proposed to explain the origin of the blue tilt \citep{mie06, bek07, str08,har09b,bla10}. 
  The self-enrichment process among these appears to be a main driver  \citep{mie06, str08}. 
We investigated any existence of this blue tilt in the M86 GCs.
Figure \ref{fig-gE3cmd} displays the CMDs of M86, M60, and M49 derived
from the $CT_1$ images for the outer regions at $1\arcmin.5 <R <7\arcmin$.
The blue and red peak values are determined from the KMM test with the colors in each magnitude bin.
In Figure \ref{fig-gE3cmd}
the blue tilt  is seen for the bright blue GCs in M86 as well as other two galaxies. 
However, the number of galaxies is only three so that it needs
to study more galaxies to address the slope difference of the blue tilt.
The existence of the blue tilt in M86 as well as M49 and M60 is consistent with the result for high mass galaxies 
in the ACSVCS given by \citet{mie10}.

The spatial distribution of the GCs in three galaxies share common features as follows:
(a) the radial number density profile of the red GCs in the outer region is consistent with the surface brightness profile of the stellar halo,
(b) the elongation of the red GC system is consistent with that of the stellar halo, 
   while the spatial distribution of the blue GC system is approximately circular,
(c) the mean color of the red GCs is similar to that of the stellar halo.
These common properties are also shown for  other gE GC systems such as
M87 and NGC 1399 \citep{bas06,tam06b,for07,fai11,str11,for12}.
Thus %
   the red GCs are more correlated with the stellar halo than the blue GCs.
This indicates that most red (metal-rich) GCs in gEs might have formed with the halo stars.

\section{SUMMARY}

We presented Washington $CT_1$ photometry of the GCs in M86 
  covering a $16\arcmin \times 16\arcmin$ field. 
We showed various photometric properties of the GC system. %
Primary results are summarized as follows.

\begin{enumerate}

 \item 
A significant population of GCs in M86 are shown in the CMD.
 
 \item
The color distribution of the GCs in M86 is bimodal,
showing a blue peak at $(C-T_1)=1.30$ and width $\sigma=0.15$, 
 and a red peak at at $(C-T_1)=1.72$ and $\sigma=0.15$,
with a division color $(C-T_1)\approx1.55$.

\item
The spatial distribution of the blue GCs is roughly circular,
while that of the red GCs is elongated similarly to that of the stellar halo.

\item
The radial number density profile of M86 GC system is more extended than the galaxy stellar light.
The radial density profile of the red GCs is consistent with the surface brightness profile of M86 stellar halo.
The radial density profile of the blue GCs is more extended than that of the red GCs. 
The density profiles at $1\arcmin<R<10\arcmin$ are approximately well fitted
by a de Vaucouleurs law and a  power law.
As a result, the slope of the radial profile of the red GCs is steeper than that of the blue GCs.
The radial density profiles in the central region are somewhat flat and 
approximately fitted by a King model.

\item
M86 GCs have the negative radial color gradient
because the number ratio of the blue GCs to the red GCs increases as galactocentric radius increases.
The mean color of the red GCs is similar to that of the stellar halo.

\item
As seen in other Virgo gEs,
the bright blue GCs in the outer region of M86 reveal a blue tilt that the mean colors of the blue GCs get redder as they get brighter.

\end{enumerate}

\acknowledgments{
The author is grateful to his collaborators: Profs. Myung Gyoon Lee and Doug Geisler.
This is supported in part
by Mid-career Researcher Program through NRF grant funded by the MEST (No.2010-0013875).
}

\end{document}